\documentclass[final,5p,times,twocolumn]{elsarticle}
\usepackage{amssymb}
\usepackage{amsmath}
\usepackage{upgreek}
\journal{Solid State Communications}

\begin{document}

\begin{frontmatter}

\title{Carrier localization in defected areas of (Cd, Mn)Te quantum well\\investigated via Optically Detected Magnetic Resonance employed in the microscale} 

\author{A. Dydniański$^{ a}$}

\affiliation{organization={Institute of Experimental Physics, Faculty of Physics, University of Warsaw},
            addressline={ul. Pasteura 5}, 
            city={Warsaw},
            postcode={02-093},
            country={Poland}}
\author{A. Łopion$^{ b, a}$}

\affiliation{organization={Institute of Physics, University of Münster},  city={Münster},postcode = {48149}, country={Germany}}

\author{M. Raczyński$^{ a}$}
\author{T. Kazimierczuk$^{ a}$}
\author{K. E. Połczyńska$^{ a}$}
\author{W. Pacuski$^{ a}$}
\author{P. Kossacki$^{ a}$}
\begin{abstract}
In this work, we study the impact of carrier localization on three quantities sensitive to carrier gas density at the micrometer scale: charged exciton (X$^{+}$) oscillator strength, local free carrier conductivity, and the Knight shift. The last two are observed in a micrometer-scale, spatially resolved optically detected magnetic resonance experiment (ODMR). On the surface of MBE-grown (Cd,Mn)Te quantum well we identify defected areas in the vicinity of dislocations. We find that these areas show a much lower conductivity signal while maintaining the same Knight shift values as the pristine areas of the quantum well. We attribute this behavior to carrier localization in the defected regions.
\end{abstract}
\end{frontmatter}
\section{\label{sec:intro} Introduction}
The optical properties of semiconductor quantum wells (QWs) with carrier gas have been studied since their first successful fabrication in the 1970s. Among many fascinating effects, the optical transitions related to neutral and charged excitons are indicators of the structure's quality and the presence of carrier gas (\cite{KKhengRCox}, \cite{KossackiPrzeglad}). Despite many years of research, a few questions still remain unanswered, one of which is the influence of carrier localization on the charged exciton spectra. In this work, we approach this problem with a technique unusual in exciton studies: optically detected magnetic resonance (ODMR). We leverage its sensitivity to the local density of spin-polarized carrier gas. The ODMR data is compared to ODMR-detected free carrier conductivity and local reflectivity spectra.

As a platform of choice for carrying out such studies we choose molecular beam epitaxy (MBE) grown (Cd,Mn)Te quantum wells with thin (tens of nanometers) (Cd,Mg)Te barriers. Their p-type doping, with the hole gas derived from the structures' surface gives rise to charged exciton observed both in photoluminescence and reflection \cite{Maslana2003_surface}. The exchange interaction between carriers and Mn$^{2+}$ ions present in said QWs is an immediate cause of giant Zeeman effect used as an optical detection method of magnetization of manganese system (\cite{Kossacki1999}, \cite{Kossacki2004}). Absorption of microwaves resonant to magnetic transitions in  Mn$^{2+}$ ions causes depolarization of their spins, which is observed as a decrease in Zeeman splitting. This in turn can be exploited for optical detection of the magnetic resonance \cite{Bogucki_strain}.

The carrier concentration in thin-barriered (Cd,Mn)Te quantum wells has been shown to be controllable by employment of additional above-barrier illumination \cite{Kossacki1999}. What is especially notable, is the realization of carrier concentration control without applying external electric field, which makes such structures an even more viable platform for studying e.g. carrier concentration effects on their magnetooptical properties.
In this paper we combine micrometer spatial movement resolution with reflectance-based ODMR measurements to map the magnetooptical properties of a (Cd,Mn)Te/(Cd,Mg)Te quantum well and investigate the effects of surface's defected areas on local charge state and properties of this heterostructure. The latter can be realized in three ways: by extracting the ratio of oscillator strengths of excitonic lines from zero-field reflectance measurements, by extracting Knight shift values from $\upmu$-ODMR measurements, which are directly proportional to carrier concentration or by analyzing ODMR background signal proportional to free carrier microwave absorption.

\section{\label{sec:sample+exp}Samples and the experimental setup}
The subject of this study was a single 10 nm (Cd,Mn)Te quantum well with Cd$_{0.77}$Mg$_{0.23}$Te barriers: 50\,nm on top of the QW and 2\,$\upmu$m on the bottom, with 3.5\,$\upmu$m CdTe buffer layer grown on SI-GaAs (100) substrate. The structure was prepared by molecular beam epitaxy. The concentration of magnesium in the barriers was determined by measurements of their bandgap energy \cite{WaagBariera}. In the previous studies on this sample the concentration of manganese in the quantum well was found to be 0.3\% \cite{Lopion2020} by fitting of the modified Brillouin function \cite{Gaj1994} to the exciton Zeeman shifts in magnetic field. Such concentration of magnetic ions is high enough to provide sufficient Zeeman splitting for well pronounced ODMR signal, but low enough so that the effects of disorder introduced by mixed material are not significant.

The magnetooptical experiments performed in this study were conducted in a cryostat equipped with superconducting coils providing magnetic fields of up to 8\,T. Two crucial for the experiment elements: microwave microstrip antenna permitting optical access to the sample (essential for ODMR measurements) and aspheric lens (NA = 0.68) mounted on $x-y-z$ piezoelectric stages, enabling micrometer spatial resolution, were integrated with the sample holder and placed along with it in the cryostat. The sample temperature in all measurements was kept at 1.7\,K (pumped helium bath) unless stated otherwise. All experiments were performed basing on reflectance measurements; the light used for accumulating spectra was emitted by a IR (770 nm) LED and it was then focused by the sample-holder-integerated lens to a spot of a size below 3\,$\upmu$m. Additional above barrier illumination used for tuning of the carrier concentration in the sample was realized by an overhead LED system, ensuring uniform and dispersed lighting.

The ODMR experiments were performed using pulsed 13.8\,GHz microwaves (MW), with duration of pulses of light and microwaves being 0.1\,ms and 3\,ms respectively. The relative timing between said pulse trains was changed to alternately record overlapped ("MW ON") and nonsimultaneous ("MW OFF") signals, and was adjusted to assure complete Mn spin relaxation in the "MW OFF" state. Such an approach was employed to ensure a fixed, constant temperature of the sample throughout the experiment.
The ODMR spectra presented furhter were obtained in a following manner: for every probed value of magnetic field, reflection spectra are taken at states "MW ON" and "MW OFF". Extracting the exciton shift values and plotting them in function of the applied magnetic field will yield ODMR spectra. One can probe the ODMR on neutral or charged exciton simultaneously.

\section{\label{sec:Zero-field}Zero-field micro-reflectance measurements}
As mentioned before, (Cd,Mn)Te/(Cd,Mg)Te quantum wells are intrinsically p-doped, with the hole gas originating from the surface states \cite{Maslana2003_surface}. Locally charged areas in the quantum well layer give rise to charged exciton, X$^{+}$, aside from the neutral exciton X. Excitonic features on optical spectra from both of these excitons can be observed in photoluminescence (PL) and in reflectance \cite{KossackiPrzeglad}. Relative intensity of the excitonic features ($A_{X^{+}}/A_{X}$) varies with changing additional above-barrier illumination and can be converted to carrier concentration. Therefore, this tuning of charge state of the sample can be employed alongside ODMR experiment, allowing to draw conclusions on charge-ion interactions by observing ODMR behavior in different charge concentration regimes in the microscale.

Spectra taken on a set point on a pristine area of the sample, with varying additional illumination are presented on Fig. 1(a). For our sample we observe that employment of additional illumination results in a decrease of carrier density evidenced by the decrease of $A_{X^{+}}/A_{X}$ ratio from 3.7 with no illumination, to 0.6 with blue (450\,nm) illumination or 0.4 under green (510\,nm) illumination. The relation $A_{X^{+}}/A_{X}$ versus carrier density $p$ is described by an empirical formula given in \cite{Kossacki1999}:
\begin{equation}
    A_{X^{+}}/A_{X} = \frac{A_0^{X^+}}{A_0^X}\frac{\sigma  p\exp{(p/p_X)}}{(2-\sigma p) \exp{(p/p_{X^+})}}
\end{equation}
where $A_0^{X^+}$ and $A_0^X$ are unscreened intensities for positively charged and neutral exciton, and $\sigma$, $p_{X^+}$ and $p_X$ are parameters describing interactions between X, X$^+$ and the hole gas. We take these parameters from \cite{Kossacki1999} and \cite{KossackiPrzeglad}.

The increase of carrier density results also in a linear increase of X-X$^{+}$ energy splitting (\cite{Kossacki2000_diss},\cite{Lopion2020}). We observe such behavior in the pristine areas of our sample (Fig. 1). Extrapolating the relation of $\Delta E_{X-X^{+}}(p)$ to vanishing carrier concentration yields charged exciton dissociation energy - in our case 2.25 meV. 

\begin{figure}[h]
    \centering
    \includegraphics{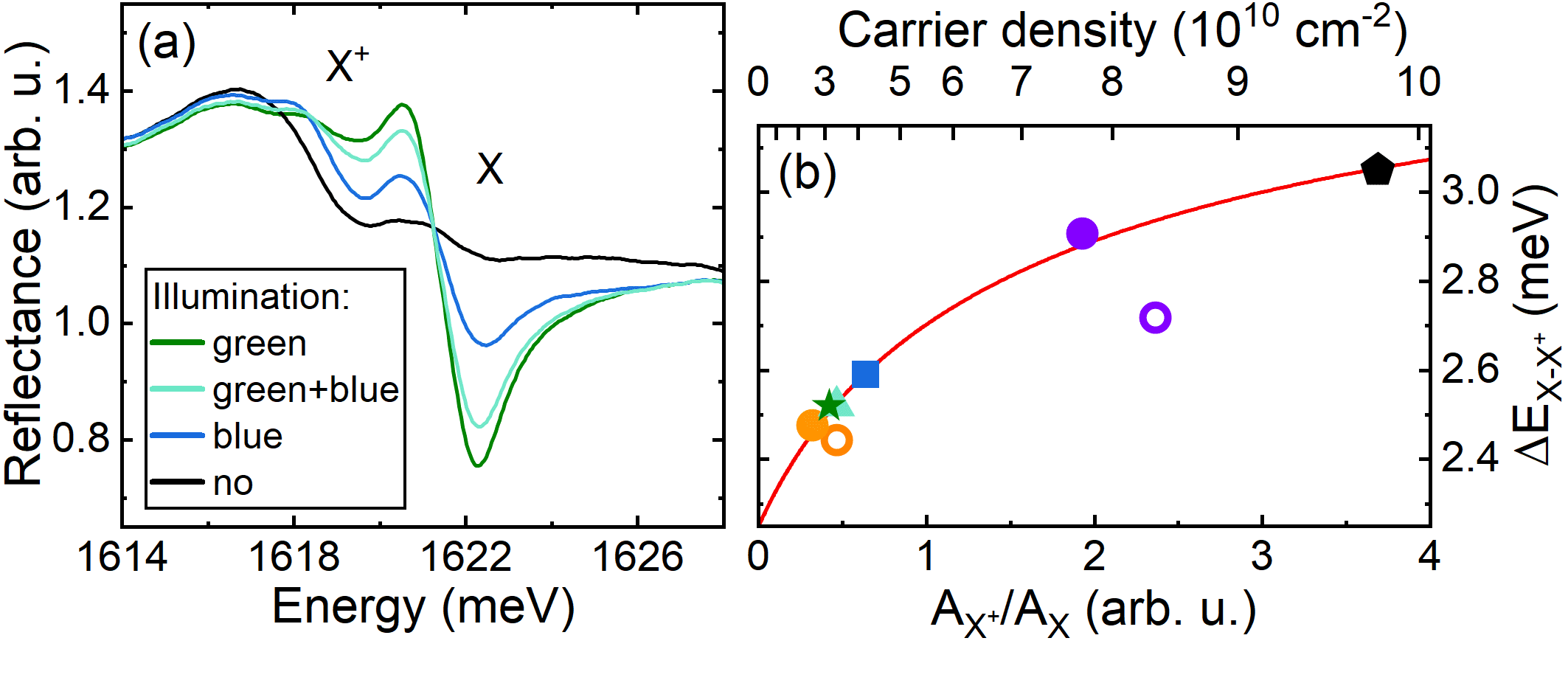}
    \caption{(a) Reflectance spectra taken at the same spot on the sample with varying above barrier illumination. (b) Energy splitting between positively charged and neutral excitons plotted in function of ratio of those excitons' amplitudes and carrier concentration. Full points - pristine areas, hollow points - defected areas. The empirical dependece and carrier density marked by the red line is taken from \cite{Kossacki1999} and \cite{KossackiPrzeglad}.}
    \label{Figure:1}
\end{figure}

In the process of scanning the sample, we have observed presence of defected areas. Single-micrometer-sized scratch-like dislocations were observed in the optical microscope image (Fig. 2(a)). Having access to precise movement resolution, we investigated the carrier concentration in proximity of such areas by performing 2D mapping of their magnetooptical properties. The mapping was realized by moving the lens on piezo stages in front of the sample and accumulating reflection spectra at every point. After the fitting process, position, amplitude and width of excitonic lines were obtained. For straightforward comparison of obtained data, a map of the $A_{X^{+}}/A_{X}$ ratio (Fig. 2(b)) and energy splitting between charged and neutral excitons $\Delta E_{X-X^{+}}$ (Fig 2(c)) is presented.

  \begin{figure}[h!]
    \centering
    \includegraphics{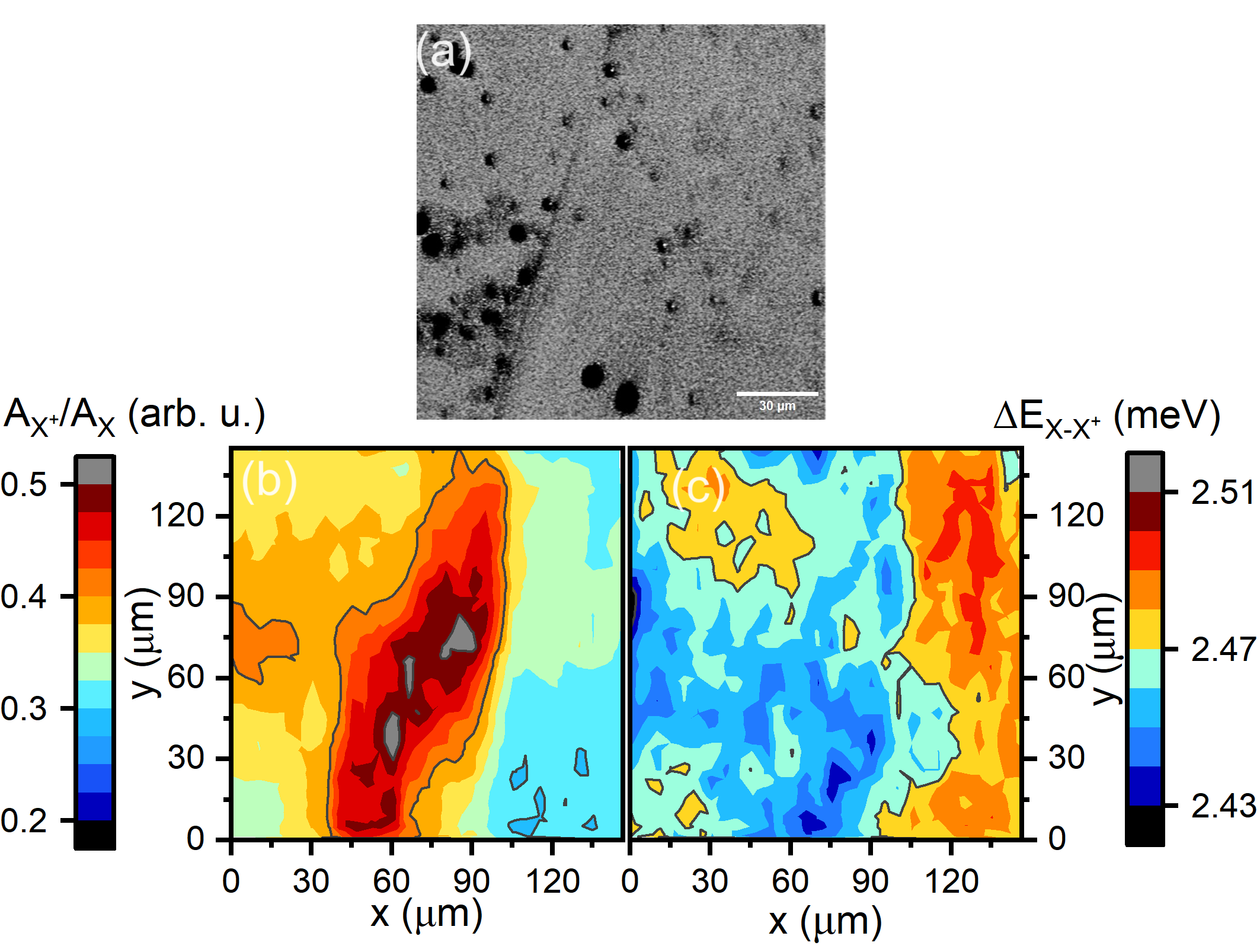}
    \caption{(a) Optical microscope photo of a defected area. (b) Map in ratio of excitonic amplitudes of the same defected region as on (a). The defected area visible as an area of high $A_{X^{+}}/A_{X}$ ratios. (c) Map of the same area but in energy splitting between neutral and charged excitons. Lower energy splitting is observed in defected areas. Maps are of size 150\,$\upmu$m x 150\,$\upmu$m, taken with a 5\,$\upmu$m step. Additional green illumination employed.}
    \label{fig:2}
\end{figure}
The defected area is unquestionably correlated with the optical properties of the sample. It exhibits stronger X$^{+}$ line (higher $A_{X^{+}}/A_{X}$) but almost unchanged or lower X-X$^{+}$ energy splitting. Such behavior clearly deviates from the tendency reported earlier for macroscopic studies and confirmed by us on a pristine region of the sample. Such a discrepancy leads us to the conclusion that the charged exciton states are significantly modified in the defected regions. This inconsistency will allow for easy identification of such regions without the need to resort to taking optical microscope images and positioning them in respect to low temperature data. The criterium for identification of discussed defected regions will be a significant deviation from the curve plotted on Fig. 1(b). Regions identified by this means have been investigated using optically detected magnetic resonance.

\section{Optically Detected Magnetic Resonance}
The ODMR experiment is a very sensitive technique to study interaction of magnetic ions with carriers in diluted magnetic semiconductor (DMS) quantum wells. They provide detailed information about magnetic ions and their close environment: crystal lattice and carriers. They have been succesfully used to study strain in the quantum well \cite{Bogucki_strain} or the temperature of the magnetic system \cite{Lopion2020}.

As the effects of illumination observed in our studies might have a complex dependence on the disorder, in the following part of our work we have decided to avoid employing any additional illumination, to probe the purest possible case of the p-type (Cd,Mn)Te quantum well.

Another defected area (different from the one on Fig. 2) that deviates from the curve on Fig. 1(b) is presented on Fig. 3(a) and 3(b). The high resolution maps were positioned so that the defected site was located on the area's left part.

\begin{figure}[h!]
    \centering
    \includegraphics{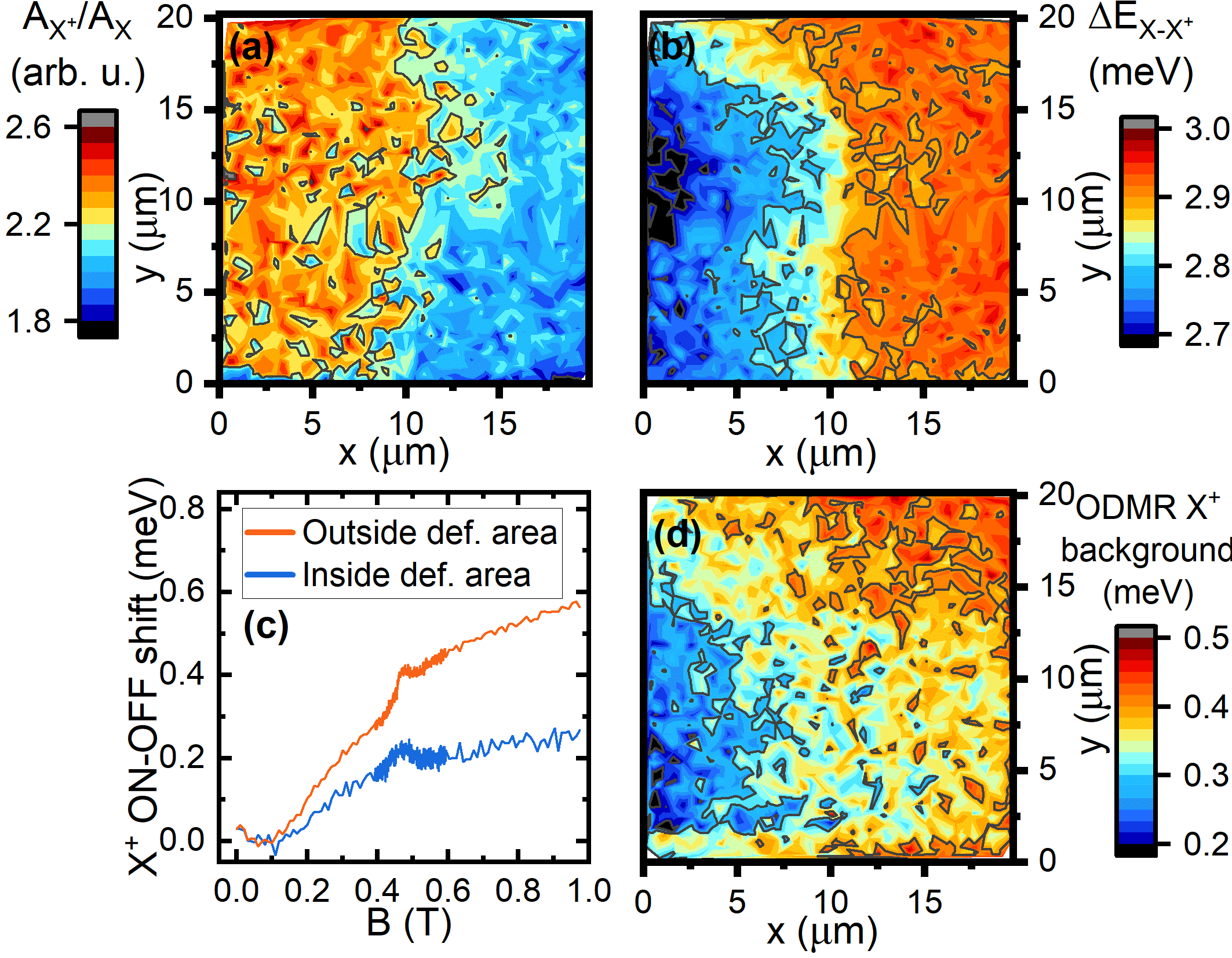}
    \caption{(a) Map in ratio of excitonic amplitudes of another area, size 20\,$\upmu$m x 20\,$\upmu$m. The defected part visible as an area of high $A_{X^{+}}/A_{X}$ ratios. (b) Map of the same area but in energy splitting between neutral and charged excitons, defected area exhibits lower energy splitting. (c) ODMR X spectra taken at points outside (orange) and inside (blue) the defected region. Non-zero ODMR background visible in both cases, with it being higher outside the defected area than inside. (d) Map of the area in ODMR background monitored on the X$^{+}$ line. Defected region seen as the part with lower ODMR background (left), and pristine area with higher background (right). B = 0.497\,T, non-resonant to X$^{+}$. All maps measured with a step 0.5\,$\upmu$m. No additional illumination employed.}
    \label{Figure:3}
\end{figure}

ODMR magnetic field scans with constant position and MW frequency were performed inside and outside of this defected region. ODMR X$^{+}$ spectra are presented on Fig. 3(c). Both spectra exhibit the manganese resonance at $\sim$0.46\,T, as well as non-resonant background. This background represents non-resonant heating of the sample by MW radiation. We attribute it to the MW absorption due to conductivity of free carriers in the quantum well. Such interpretation is supported by the spatial nonuniformity correlated with other optical properties of the QW. The significant electric field contribution of MW radiation characteristic for the microstrip antenna additionally explains such effects. We make use of this background to get information on the local conductivity of the quantum well.

In the case of spectra presented on Fig. 3(c) the conductivity is lower inside the defected area than in the pristine part. To confirm that these findings were correlated with the defected region's appearance, a map of ODMR background of this area was measured, and is shown on Fig 3(d). The background map exhibits a similar shape to Fig. 3(a) and 3(b) and there is no doubt that its correlated with the defected region's appearance. The results are in line with the conclusions drawn from the ODMR spectra - the area covered by the defected region is exhibiting low ODMR background, whereas the pristine area displays ODMR background approximately 2 times as high.

It is worth noting that in low carrier concentration regime (for example with additional green illumination) such ODMR background is not observed or is on the noise level. It is in agreement with our interpretation of the ODMR background observed both outside and inside the defected area being caused by high intrinsic carrier density.

\section{ODMR Knight shift}
The carrier-density-induced shift in magnetic field of Mn$^{2+}$ resonance in DMS systems is known in the literature as the Knight shift \cite{StoryKnight}. In the case of ODMR experiment, this can be probed as the shift between resonances in ODMR X and ODMR X$^{+}$ \cite{LopionArxiv}. In studied (Cd,Mn)Te/(Cd,Mg)Te quantum wells the positively charged exciton arises from the interaction of the exciton with the two-dimensional hole gas - in this case the ODMR X$^{+}$ resonance curve will be shifted into lower magnetic fields compared to ODMR X. For fully polarized hole gas we can approximate:
\begin{equation}
B_{\textrm{eff}} \approx \frac{-\beta p}{2g_{\textrm{Mn}}\mu_{\textrm{B}}}\frac{1}{d}
\end{equation}
with $B_{\textrm{eff}}$ being the Knight shift, $p$ being the surface hole density, $\beta$ denoting the p-d exchange constant, $g_{\textrm{Mn}}$ = 2, $\mu_{\textrm{B}}$ being the Bohr magneton and $d$ the width of the QW. The Knight shift values are directly proportional to the hole  density, which adds yet another possibility of studying carriers in (Cd,Mn)Te quantum wells. 

ODMR spectra were measured inside and outside the defected area, and the shift between resonances of ODMR X and ODMR X$^{+}$ was compared. The results (with the background subtracted for clarity) are shown on Fig. 4.

\begin{figure}[h!]
    \centering
    \includegraphics{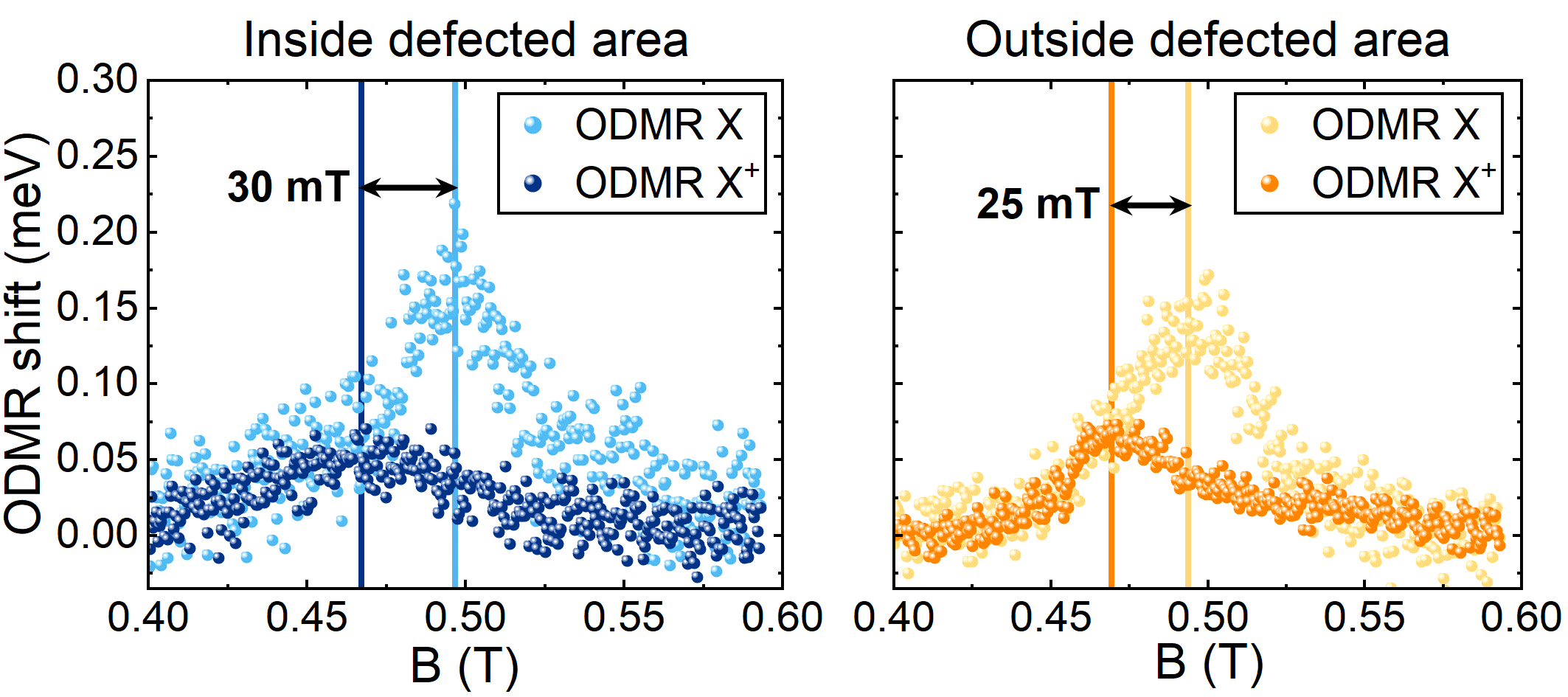}
    \caption{ODMR on X and X$^{+}$ spectra with subtracted background as a comparison of ODMR Knight shifts. Measurements taken inside the defected area (left panel, blue), and outside of it (right panel, orange). Little difference is seen between the Knight shift values in these areas (30 mT vs  25 mT) which allows for the approximation of carrier concentration being constant across the sample.}
    \label{Figure:4}
\end{figure}
The Knight shift values determined from these measurements were similar: 30\,mT inside and 25\,mT outside the defected region. Employing Eq. 2 on these results yields carrier concentration values of respectively 1.2$\times10^{11}$\,cm$^{-2}$ and 1.0$\times10^{11}$\,cm$^{-2}$. The differences between them are too small to consider them significant. Therefore the outcome of these measurements is clear - the concentration of carriers weakly (or not at all) depends on the micro-sized defected areas present on the surface of studied (Cd,Mn)Te quantum wells and is approximately constant across the sample.\\
Combining the results developed in this study leads to a single conclusion - studied defected areas indeed cause localization of carriers. The low ODMR background inside the defected region, combined with similar Knight shift values inside and outside of the defected area shows, that simple variation of the carrier density is not sufficient to explain observed effects. The defected region apparently shows lower density of $free$ carriers, with the overall carrier concentration remaining unchanged. Additionally, localized carriers cause decrease of X-X$^{+}$ energy splitting and increase in X$^{+}$ intensity in absorption.

\section{Summary}
In this study we investigated single-micrometer sized defected areas found on the surface of single 10nm (Cd,Mn)Te/(Cd,Mg)Te quantum wells and their impact on the magnetooptical properties of studied samples. We characterized the carrier concentration optically by three means - by determining the oscilator strength from the reflection spectra and calculating the carrier concentration from the previously developed empirical model, and by measurements of Knight shifts of optically detected magnetic resonance employed in the microscale ($\upmu$-ODMR). We had also investigated the origin of ODMR background present in a wide range of magnetic fields and came to a conclusion that it is caused by non-resonant free carrier absorption, which surprisingly turned out to be lower inside than outside of the studied defected areas. Combining this fact with almost constant carrier concentration across the sample (determined from ODMR Knight shifts) we propose that studied defected areas cause localizations of carriers and are not just less/more locally charged areas. At the same time we prove $\upmu$-ODMR to be a valid technique of optical determination of carrier concentration and carrier interactions with topologically-distinct areas on the sample's surface.

\section{Acknowledgements}
This work was supported by the Polish National Science Centre under decisions DEC-2020/39/B/ST3/03251, DEC-2020/38/E/ST3/00364, and DEC-2021/41/B/ST3/04183.

\bibliographystyle{elsarticle-num-names} 
\bibliography{A_Dydnianski_bibliography}
\end{document}